\documentclass[prl,floatfix,twocolumn,preprintnumbers,amsmath,amssymb,superscriptaddress]{revtex4}
\usepackage{color}

\usepackage{graphicx,psfrag} 
\usepackage{hyperref}
\usepackage{verbatim}
\usepackage[utf8]{inputenc}

\usepackage{soul}

\newcommand{\newc}{\newcommand}
\newc{\gsim}{\lower.7ex\hbox{$\;\stackrel{\textstyle>}{\sim}\;$}}
\newc{\lsim}{\lower.7ex\hbox{$\;\stackrel{\textstyle<}{\sim}\;$}}
\newc{\gev}{\,{\rm GeV}}
\newc{\mev}{\,{\rm MeV}}
\newc{\ev}{\,{\rm eV}}
\newc{\kev}{\,{\rm keV}}
\newc{\tev}{\,{\rm TeV}}

\definecolor{mypurple}{RGB}{164,64,214}

\usepackage{amsmath}
\def\beq{\begin{equation}}
\def\eeq{\end{equation}}
\def\bea{\begin{eqnarray}}
\def\eea{\end{eqnarray}}
\def\bitem{\begin{itemize}}
\def\eitem{\end{itemize}}
\newc{\ie}{{i.e. }}          \newc{\etal}{{\it et al.}}
\newc{\eg}{{e.g. }}          \newc{\etc}{{\it etc.}}
\newc{\cf}{{\it c.f.}}
\def\bar#1{\overline{#1}}

\def\inv{^{\raise.15ex\hbox{${\scriptscriptstyle -}$}\kern-.05em 1}}
\def\lbar{{\lower.35ex\hbox{$\mathchar'26$}\mkern-10mu\lambda}} 

\let\<=\langle
\let\>=\rangle

\let\+=\uparrow

\begin{document} 
\title{The String Soundscape at Gravitational Wave Detectors}
\author{Isabel Garc\'ia Garc\'ia}
\email{isabel.garciagarcia@physics.ox.ac.uk}
\affiliation{Rudolf Peierls Centre for Theoretical Physics, University of Oxford,
1 Keble Road, Oxford OX1 3NP, UK}
\author{Sven Krippendorf}
\email{sven.krippendorf@physics.ox.ac.uk}
\affiliation{Rudolf Peierls Centre for Theoretical Physics, University of Oxford,
1 Keble Road, Oxford OX1 3NP, UK}
\author{John March-Russell}  
\email{jmr@thphys.ox.ac.uk}  
\affiliation{Rudolf Peierls Centre for Theoretical Physics, University of Oxford,
1 Keble Road, Oxford OX1 3NP, UK}

\begin{abstract}
We argue that gravitational wave (GW) signals due to collisions of ultra-relativistic bubble walls may be common
in string theory.
This occurs due to a process of post-inflationary vacuum decay via quantum tunnelling within (Randall-Sundrum-like) warped
throats.   Though a specific example is
studied in the context of type IIB string theory, we argue that our conclusions are
likely more general.  Many such transitions could have occurred in the post-inflationary Universe, as a large number of throats with
exponentially different IR scales can be present in the string landscape, potentially leading to several
signals of widely different frequencies -- a \emph{soundscape} connected to the landscape of vacua.  Detectors such as eLISA and AEGIS, and observations with BBO, SKA and EPTA (pulsar timing) have the sensitivity to detect such signals, while at higher frequency aLIGO is not yet at
the required sensitivity.  A distribution of primordial black holes is also a likely
consequence, though reliable estimates of masses and $\Omega_{\rm pBH} h^2$ require dedicated numerical simulations, as
do the fine details of the GW spectrum due to the unusual nature of both the bubble walls and transition.
\end{abstract}

\maketitle 
\section{\label{intro}I. Introduction}

The recent direct detection of gravitational waves (GW) by LIGO \cite{Abbott:2016blz} opens a new mode of physical exploration.
Although the potential of GW detectors to study astrophysical objects has
been deeply investigated \cite{1511.05994}, their potential for exploring Beyond-the-Standard-Model (BSM) physics is still in a relative infancy.
Prime examples studied include the physics of inflation~\cite{Grishchuk:1974ny,Starobinsky:1979ty,Rubakov:1982df,Fabbri:1983us,Abbott:1984fp}, the presence of strongly first order thermal phase transitions in the early Universe, e.g.~non-SM electroweak (see~\cite{Witten:1984rs,Turner:1992tz,Kosowsky:1992rz,Kosowsky:1991ua,astro-ph/9211004,astro-ph/9310044} for early work and~\cite{Caprini:2015zlo} for a recent overview), and the possibility of probing the existence of axions~\cite{Arvanitaki:2016qwi}.

\begin{figure}
  \includegraphics[scale=0.35]{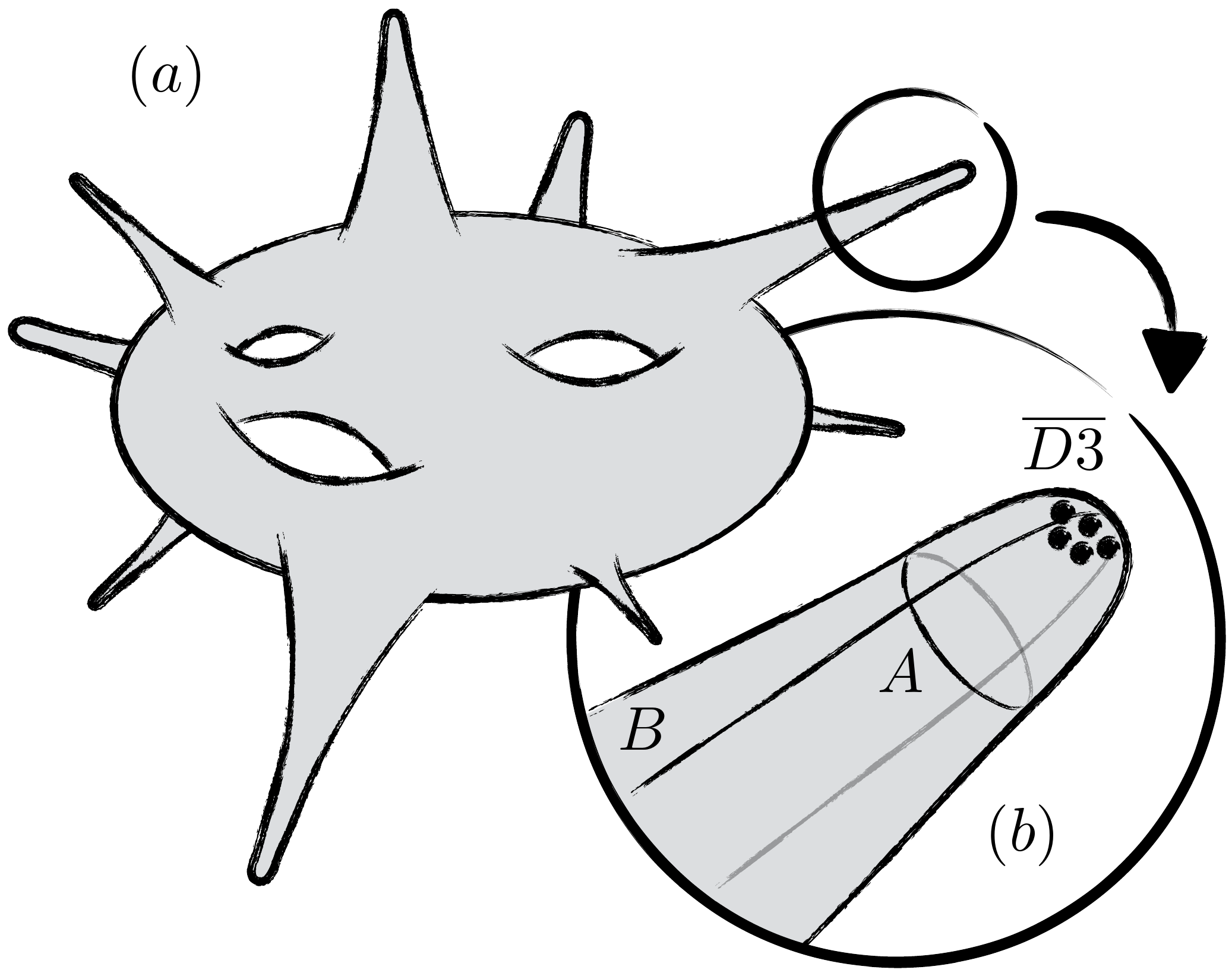}
\caption{\label{fig:cartoon}
	(a) Cartoon of a type IIB flux compactification featuring a large number of warped regions (throats) 
	some of which will be of Klebanov-Strassler (KS) type \cite{Klebanov:2000hb}.
	(b) Close-up of a KS throat (topologically $\cong S^3 \times S^2 \times \mathbb{R}$) with 3-form RR and NSNS flux quanta on the $A$-cycle and on the $B$-cycle respectively.
	The fluxes lead to a tip warp factor $w_{IR}$.  In the locally SUSY-breaking false vacuum anti-$D3$ branes are localized at the tip~\cite{Kachru:2002gs}.}
\end{figure}

We argue that GW detectors may provide a powerful tool to interrogate the nature of short-distance physics, particularly string theory, in a way unrelated to the process of inflation: specifically, GW signals from post-inflationary vacuum decay are a natural feature of the type IIB (and likely more general) string landscape.  Our conclusions rely on the observation that flux compactifications in type IIB string theory can contain a large number of highly warped regions~\cite{Giddings:2001yu,Douglas:2003um,hep-th/0307049,manythroats1,manythroats2}, often referred to as throats, with physics related to that of Randall-Sundrum models~\cite{hep-ph/9905221,hep-th/9906064} (see Fig.~\ref{fig:cartoon}).
Under rather general assumptions, later made more precise, a throat can present a metastable vacuum in which supersymmetry (SUSY) is locally broken, along with a true locally-SUSY-preserving vacuum \cite{Kachru:2002gs}, to which it eventually decays.

Specifically, we explore the process of vacuum decay,
taking place within a given throat, via the process of zero-temperature quantum tunnelling.
We study the effect of the nucleation of bubbles of the true vacuum in the early Universe, and argue that resulting
ultra-relativistic bubble wall collisions may lead to observable GW signals.
The frequency of the GW produced will be different for different throats, since it depends sensitively on its details, most of all
on the gravitational warp factor (i.e. red-shift), $w_{IR} \ll 1$, setting the relation between the IR energy scale of the tip of the throat and the string
scale~$M_{s}$. Since a large number of throats with exponentially different warp factors can be present in the string landscape
(see e.g.~\cite{Hebecker:2006bn}), GW signals with very different frequencies can be produced.

Space-based experiments, such as BBO \cite{BBO1,BBO2}, eLISA \cite{eLISA}, or AEGIS \cite{AEGIS3},
and pulsar timing arrays like SKA \cite{SKA} or EPTA \cite{EPTA}, are well suited for detection of these \emph{soundscape} signals, both in terms of frequency range and sensitivity.
Larger compactification volumes and smaller $w_{IR}$ both shift the frequency peak of the signal towards smaller values, making pulsar timing array detectors optimal for probing very large volume and/or very strongly warped scenarios, while
in the higher frequency range where aLIGO \cite{aLIGO} operates, even the strongest GW signal is unlikely to be detectable by current technology.

Another likely consequence of the ultra-relativistic bubble wall collisions is the production
of primordial black holes (pBHs)~\cite{PBH1,PBH2,PBH3,PBH5,PBH6}.  This pBH production process, and the fine details of the high-frequency portion
of the GW spectrum itself, is sensitive to the peculiarities of the bubble wall and vacuum
decay dynamics that apply in our case
(the dynamics are different than those of both thermal phase transitions,
and previous studies of inflation-terminating quantum tunnelling vacuum decay).
A reliable calculation of 
the pBH mass distribution and of $\Omega_{\rm pBH} h^2$ requires dedicated numerical simulations, as does the GW spectrum in the
frequency domain beyond the peak position.  If the production of pBHs is both highly efficient, and has a mass distribution that extends
above $\simeq 10^9 \ {\rm g}$, then the maximum amplitude of GWs observable today can be constrained.  On the other hand,
the possible production of pBHs in the mass regions where pBHs may still comprise a significant fraction of the dark matter (DM) density provides
another motivation for detailed studies.   We emphasise that the study we present here is just a first step
towards understanding the rich physics of the string soundscape.

\section{II. False Vacuum Decay}

{\bf Outline of early Universe history -- }
We assume that after inflation ends, the visible sector (i.e. the Standard Model (SM) plus any other states in significant thermal contact) gets reheated to some temperature $T_{rh}$.  We conservatively take $T_{rh} \gtrsim 4 \ {\rm MeV}$ so that Big Bang nucleosynthesis (BBN) can proceed undisrupted.
(This assumption may be relaxed within more general early Universe histories.)
But other sectors, such as those living at the end of highly warped throats, may not be reheated to the same degree and, in general,
one would expect many, though possibly not all, of those hidden sectors being left at temperatures $T \ll T_{rh}$.  This is a natural assumption
given that we do not observe a much larger DM-to-baryon ratio, nor a significant number of excess relativistic degrees
of freedom, $\Delta N_{eff}$.   Thus in the following we take, for simplicity, the throat sector under consideration to be at temperature $T_{th} = 0$.
(Strictly, all we require is that $T_{th}$ is much smaller than all mass scales present in the problem,
in which case a $T_{th} = 0$ treatment suffices.
The case $T_{th}\neq 0$ may lead to a \emph{thermally assisted} quantum tunnelling
decay, or a \emph{purely thermal transition} if $T_{th}$ is high enough, similar to the Randall-Sundrum case \cite{Creminelli:2001th,Hassanain:2007js,Konstandin:2010cd}.
These possibilities are studied in upcoming work \cite{future1}.)
This is a self-consistent assumption since the IR dynamics of a throat are known to be sufficiently (though not absolutely) sequestered from the dynamics
of the rest of the compactification \cite{Halter:2009nb}.

In this set-up, the throat sector remains in a metastable vacuum
so long as the probability of nucleating a bubble of true vacuum in a Hubble
volume in a Hubble time is much smaller than one, i.e.~$\Gamma / H(T)^4 \ll 1$,
where $\Gamma$ is the decay rate per unit volume (independent of temperature since $T_{th} = 0$)
and $H(T)$ is the Hubble rate, dependent on the \emph{visible} sector temperature $T$.
Only when $\Gamma / H(T)^4 \approx 1$ does the decay occur.
(We remark that as the decay takes place when $\Gamma / H(T)^4 \approx 1$, one does not need to worry about counting
of negative fluctuation modes of the bounce solution \cite{Coleman:1987rm}, since in this regime it is proven
that one and only one negative
mode is present and therefore the Euclidean bounce solution is guaranteed to correctly compute the false vacuum decay~\cite{Lee:2014uza}.
This feature is not assured in the much different parameter regime
considered in \cite{Kachru:2002gs}, so it is not clear if in \cite{Kachru:2002gs} a \emph{physical} decay rate has been calculated or not.)
Throughout, we will assume that the radiation energy density of the visible sector, $\rho_{rad}(T)$, dominates the false vacuum energy density of the  throat 
\footnote{The case in which a phase of matter domination takes place either during of after the transition, but before the start of BBN, would yield different results
to those presented in this letter.  A period of matter domination is a well motivated possibility (see~\cite{Kane:2015jia} for a recent overview), and leads to interesting variant phenomenology, \cite{future1}.}.
Defining
\begin{equation}
	\alpha (T) \equiv \frac{\rho_{vac}}{\rho_{rad} (T)}~,
\end{equation}
we thus take $\alpha (T) \lesssim 1$ for all temperatures of interest.
This ensures that the Universe is always radiation dominated (RD), so that a second inflationary phase never takes place
and phenomenologically disastrous, large late time density perturbations do not occur \cite{Guth:1982pn, La:1989st}.
Notice that since $\rho_{rad} (T) \sim T^4$, $\alpha (T)$ decreases as the Universe expands, so even if at the time of bubble wall
collision $\alpha_c \equiv \alpha(T_c) \approx 1$ ($\alpha_c$ will set the strength of the GW signal),
at the epoch of bubble nucleation $\alpha_n \equiv \alpha(T_n) < 1$.

Both the nucleation temperature $T_n$ (set by $\Gamma$), and the bubble properties, depend on the microphysics of our specific
type IIB model to which we now briefly turn.  Readers primarily interested in the resulting GW phenomenology may jump ahead to
the next sub-section. 

{\bf Metastable throats --}
The authors (KPV) of \cite{Kachru:2002gs} considered the dynamics of $p$ anti-$D3$ branes ($\bar{D3}$) in a Klebanov-Strassler (KS) throat \cite{Klebanov:2000hb} (see Fig.~\ref{fig:cartoon}).
In this conifold geometry (topologically $\cong S^3 \times S^2 \times \mathbb{R}$), $M$ units of RR 3-form flux pierce the $A$-cycle ($\cong S^3$ of the conifold), whereas $K$ units of NSNS 3-form flux pierce the dual $B$-cycle which extends in the non-compact directions in the local model (i.e.~into the bulk of the geometry when embedded into a compact manifold).  These fluxes result in a tip warp factor $w_{IR} \sim \exp(-2\pi K/3M g_s)$ where $g_s$ is the string coupling~\cite{Giddings:2001yu}.
Ignoring both back-reaction local to the throat, and back-reaction arising from other distant parts of the compactification (we later comment on these issues), in \cite{Kachru:2002gs} it was argued that if the ratio $p/M$ was smaller than a certain critical value $r_c = (\pi-3+b_0^4)/(4 \pi) \approx 0.08$
then the system features a metastable vacuum in which SUSY is locally broken by the $\bar{D3}$-branes.  Decay to the true SUSY-preserving vacuum, with no $\bar{D3}$-branes, $(M-p)$ $D3$-branes, and $(K-1)$ NSNS flux quanta could only take place quantum mechanically, or through a thermal transition.

The fluxes are constrained by a `tadpole condition', a generalisation of Gauss' law that relates, in a 6-dimensional Calabi-Yau (CY) manifold,
the tadpole $L$ (depending on the topology of the CY, and other types of $D$-branes and $O$-planes),
the number of $D3$ branes and the RR and NSNS background fluxes: $L = N_{D3} - N_{{\bar{D3}}} + h_i \Sigma_{ij} f_j$, where $\Sigma$ is the symplectic matrix.
The individual fluxes, $\{ h_i, f_i \}$, and associated flux quanta $\{ K^{i}, M^{i} \}$ that can be chosen depends on the number of moduli in the CY.
Typical known CYs feature up to thousands of moduli (see for instance~\cite{Kreuzer:2000xy}), and lead to a tadpole $L$ of similar sizes.
Thus individual flux choices such as $K, M$ are constrained to be at most of the same order as $L$. However this constraint leaves still an enormous number of possible flux configurations, and motivates our choice of `typical' fluxes $K,M\sim{\mathcal O}(10^2)$.  In addition, the string coupling, $g_s$, is stabilised supersymmetrically by a function of these flux choices. To study a type~IIB effective action (as we do here) requires a string coupling $g_s\ll 1$.  The correct SM gauge couplings can easily be accommodated with $g_s\sim\mathcal{O}(10^{-2})$~\cite{Cicoli:2013mpa}.

In the metastable vacuum, the system is not well described in terms of individual $\bar{D3}$ branes, but rather as an NS5 brane.
This NS5 brane (a 5-dimensional object) has 3 non-compact spatial dimensions, the remaining 2 being wrapped around an $S^2$ contained in the $S^3$
of the conifold geometry.  The position of this $S^2$ within the $S^3$ is described by an angular variable $\psi$.
The state of the system can then be encapsulated by the dynamics of a scalar field $\psi$, initially in a false vacuum $\psi_{fv} \in [ 0, \pi/4 )$,
and whose value in the true vacuum is $\psi_{tv} = \pi$ (where the radius of the NS5 brane is zero).
The Lagrangian describing this system (setting $M_{s} \equiv 1/\sqrt{\alpha'}=1$ and in red-shifted units, so hiding the warp factor $w_{IR}$) is
\cite{Kachru:2002gs}
\begin{equation}\nonumber
	\mathcal{L} \approx \frac{\mu_3 M}{g_s} \left( - V_2(\psi) \sqrt{1 - \partial_\mu \psi \partial^\mu \psi} + \frac{1}{2 \pi} (2 \psi - \sin 2\psi) \right),
\label{eq:lagrangian}
\end{equation}
\begin{equation}
	V_2(\psi) = \frac{1}{\pi} \sqrt{ b_0^4 \sin^4\psi + (\pi \frac{p}{M} - \psi + \frac{1}{2} \sin 2 \psi)^2 }~,
\end{equation}
with $\mu_3 = (2 \pi)^{-3}$ and $b_0^2 \approx 0.93266$.  When $p/M< r_c$ the potential of this
system has a local minimum below $\psi=\pi/4$, while for $p/M \geq r_c$ there exists only the minimum at $\psi_{tv}=\pi$. 
(We refer the reader to \cite{Kachru:2002gs} for further details.)

Note that the local non-compact KPV set-up, used in this letter, suffers from back-reaction of $\bar{D3}$-branes on the geometry.
This issue has led to a long-standing debate~(see \cite{hep-th/0402088,0910.4581,0912.3519,1202.1132,1410.7776,1410.8476,1411.1121,1412.5702,1502.01234, 1502.07627,1507.07556,Polchinski:2015bea} for a selection of issues), but as of now there is no definitive full string theory calculation on this issue.  We work under the assumption that the EFT point of view in~\cite{Polchinski:2015bea} is appropriate, and hence that the local back-reaction will not change the \emph{qualitative} features of this system but only small quantitative changes will occur (likely resulting in a somewhat smaller value of $r_c$).

{\bf  Bubble nucleation --}
KPV~\cite{Kachru:2002gs}, considered the case $p/M \ll r_c$ where the false vacuum decay rate (\emph{if} a decay)
leads to a lifetime $\tau\ggg 10^{10} {\rm yrs}$.
We instead focus on the case $p/M \lesssim r_c$, close to the regime of classical instability. We thus introduce $\delta$ defined as
\begin{equation}
	\frac{p}{M} \equiv r_c (1 - \delta) \qquad \qquad 0 < \delta \ll 1~.
\end{equation}
We stress that our consideration of one or more throats close to classical instability is not unreasonable: given the large number of throats
typically present in the type~IIB string landscape, \emph{some} of them can find themselves in this near-to-critical situation.  (We return to this below.)

\begin{figure}
  \includegraphics[scale=0.53]{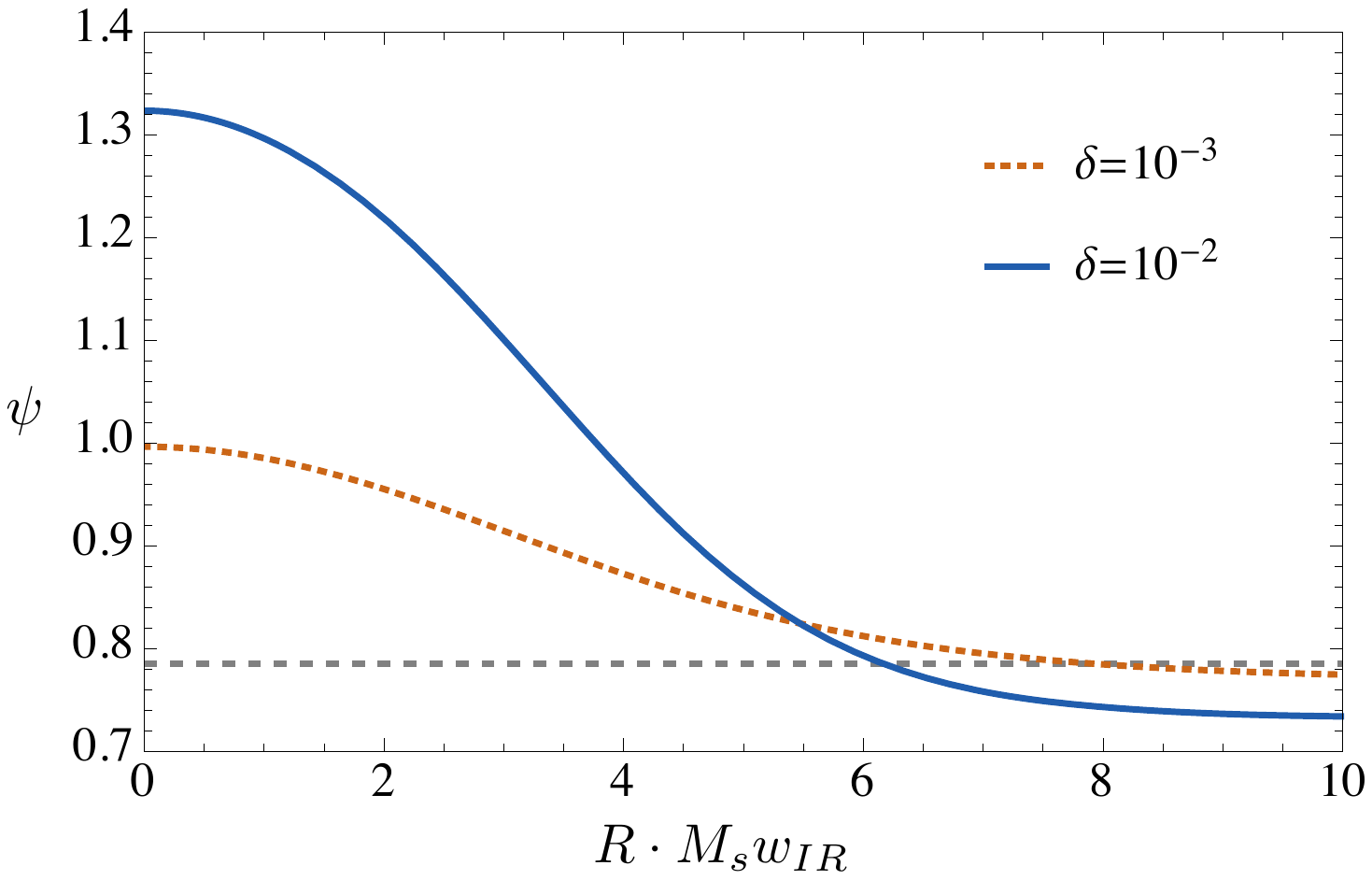}
\caption{\label{fig:bubbleprofile}
	Profile of a critical bubble at nucleation as a function of $\tilde R \equiv R \cdot M_s w_{IR}$,
	for two different values of $\delta$. In both cases, a thick-wall profile occurs.
	The dashed line is $\psi = \pi / 4$, the asymptotic value,  as $\delta \rightarrow 0$, of the field
	in the false vacuum. } 
\end{figure}

Following Coleman \cite{Coleman:1977py}, the decay rate per unit volume can be written as $\Gamma \sim m^4 e^{-B}$,
where $B$ is defined as $B = S[\psi_B] - S[\psi_{fv}]$, where $\psi_B$ is the field configuration that extremises the Euclidean action
(the \emph{bounce} solution), and $\psi_{fv}$ refers to the static false vacuum configuration. The pre-factor results from the (square-root) determinant of the quadratic fluctuation operator around the bounce, the associated mass scale $m$ being sufficiently well approximated by
the curvature around the barrier.  We find (restoring powers of $M_s$ and $w_{IR}$)
\begin{equation}
	\left( \frac{m}{M_s} \right)^4 \approx \frac{2^{11} \pi  b_0^8 r_c}{(1 + b_0^4)^5} w_{IR}^4  \delta
						\approx 17 w_{IR}^4  \delta~.
\end{equation}
Moreover, since the metastable vacuum is close to classical instability, the vacuum energy density, $\rho_{vac}$,
is much larger than the barrier height, and thus one should expect the bubbles at nucleation to show a thick-wall profile
(rather than a thin-wall profile as considered in \cite{Kachru:2002gs} for the case $p/M \approx 0$).
In this regime, $\psi_B(\tilde R)$ ($\tilde R \equiv R \cdot M_s w_{IR}$, with $R$ being the SO(4)-invariant Euclidean radius) will change slowly with $\tilde R$, an observation that allows
the bounce equation to be simplified as
\begin{equation}
	\frac{d^2 \psi}{d^2 \tilde R} + \frac{3}{\tilde R} \frac{d \psi}{d \tilde R} \approx \frac{\pi V_2(\psi)^\prime - 1 + \cos(2 \psi)}{\pi V_2(\psi)} ~,
	\label{eq:bounce}
\end{equation}
which can be solved numerically using the undershoot/overshoot method (we utilise \cite{Wainwright:2011kj}).
(In Fig.~\ref{fig:bubbleprofile} we show the critical bubble profile for two different values of~$\delta$.  As expected, the bubbles
show a thick-wall profile at the time of nucleation, and the value of $\psi$ at the centre of the bubble is initially well below the true
vacuum value $\psi_{tv} = \pi$.)
Once the bounce solution $\psi_B({\tilde R})$ has been obtained, numerical evaluation of the bounce action $B$ becomes straightforward.
We obtain
\begin{equation}
	B  = 2 \pi^2 \mu_3 b_0^4 g_s M^3 f(\delta)
		 \approx 36 \ \frac{g_s}{0.03} \left( \frac{M}{10^2} \right)^3 \frac{f(\delta)}{f(10^{-3})}
\end{equation}
where $f(\delta) \approx 0.38 \ \delta^{1/2} + 6.0 \ \delta$
(for $\delta \approx 10^{-4}$ to $10^{-2}$).

Some comments are in order.  The reader might be worried that gravitational effects may be important when studying the
decay process \cite{Coleman:1980aw}. This is not the case, as in the region of parameter space we consider the radius of the bubbles at the time of nucleation (the critical radius $R_c$) always satisfies $R_c \ll H(T_n)^{-1}$.
Secondly, one might question our entitlement to vary $\delta$ essentially continuously: since $p$ refers to the number of $\bar{D3}$ branes at the tip of the throat, and $M$ to the units of RR flux, one would think that only discrete values of their ratio, and thus discrete values of $\delta$, can be considered.
However, small, but important corrections appear when embedding the local KPV setup we utilise into a complete, global compactification manifold depending on the suppression of couplings between local and global modes (sequestering)~\cite{Kachru:2003sx,Baumann:2006th}.
For us, this dependence of local parameters (e.g.~$w_{IR}$, and especially $\delta$) on bulk properties, including the enormously large number of distant flux
values $\{K^i, M^i\}$, effectively produces a very finely grained discretum \cite{ArkaniHamed:2005yv}, justifying our choice of varying an \emph{apriori} discrete parameter $\delta$ continuously.  

\section{III. Gravitational  Waves}

{\bf Expansion of bubbles --}
At nucleation, the bubble walls are spherical and at rest
\footnote{Both these statements are approximations up
to corrections of order $1/\sqrt{B}$ \cite{Garriga:2013pga,Bond:2015zfa}, which is an $SO(2,1)$ symmetry breaking effect.
As a result, a full treatment of the high-frequency part of the GW spectrum, and the details of pBH production require inclusion of these effects.}.
After a time $\sim R_c$ they are expanding with relativistic velocities and, because of the absence of a
thermal plasma with which they meaningfully interact, they continue to accelerate to ultra-relativistic velocities.
Since we consider a zero-temperature tunnelling process, production of gravitational radiation from the decay arises solely from collision of bubbles:
effects like sound-waves or turbulence in the thermal plasma, which modify the GW spectrum in the case of thermal transitions \cite{Caprini:2015zlo}, are, to a good approximation, not present.  Moreover, unlike in most thermal phase transitions where the temperatures at bubble nucleation and bubble collision are very close to each other, in our situation this is not the case. 
Bubbles of critical size are nucleated at temperature $T_n$ (or time $t_n$), when $\Gamma/H(t_n)^4 \approx 1$.
Since the radius of this critical bubbles is $R_c \ll H(t_n)^{-1}$, we can treat them as point-like,
and therefore the average separation between nearest bubbles is $d(t_n) \sim H(t_n)^{-1}$.
so after a time $\Delta t \gg R_c$
their radius is $R(t) \sim \Delta t$.
On the other hand, the centre of two bubbles has expanded further apart in the way that corresponds to a RD Universe
$d(t_n + \Delta t) \sim H(t_n)^{-1} a(t_n + \Delta t)/a(t_n) = 2 \sqrt{t_n (t_n + \Delta t)}$
and so the bubble walls finally collide at a time $\Delta t \approx 1.6 t_n$ after nucleation. This translates into a temperature at collision,
$T_c = T_n \sqrt{t_n / t_c} \approx T_n/\sqrt{2.6} \approx 0.62 \ T_n$.
This distinction between $T_n$ and $T_c$ becomes important for the GW spectrum, since it is $T_c$ that will set the
frequency peak of the corresponding GW signal.
Notice that $T_c \approx 0.62 T_n$ leads to $\alpha_n \approx \alpha_c (T_c/T_n)^4 \approx 0.15 \alpha_c < \alpha_c$.

\begin{figure}
  \includegraphics[scale=0.53]{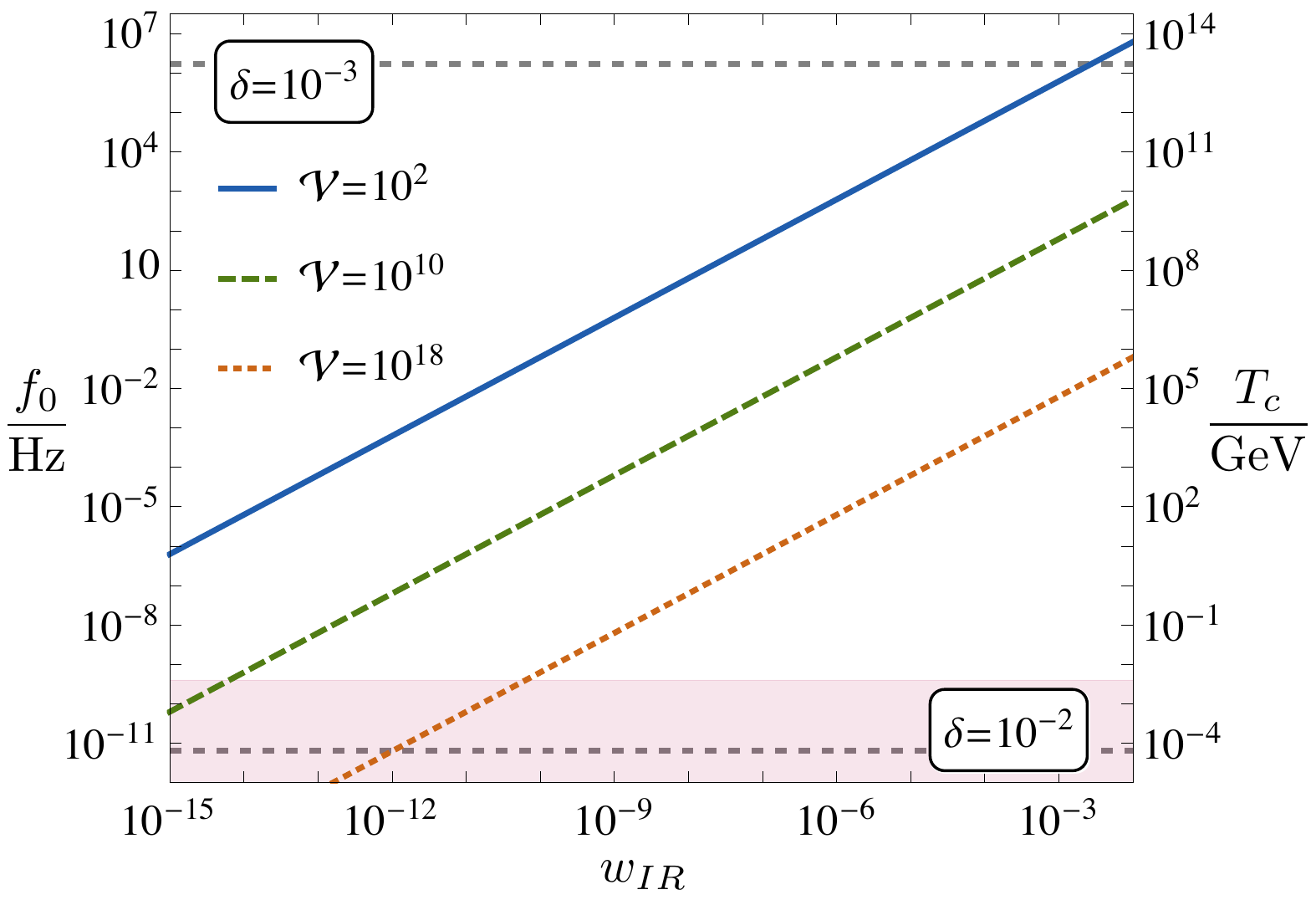}
\caption{\label{fig:freqvsepsilon}
	Approximate value of the peak frequency of the GW signal
	as a function of the warp factor, $w_{IR}$, for 3 different values of the volume of the 6-dimensional manifold.
	For illustration, we take $M=10^2$ and $g_s = 0.03$. 
	The shaded pink region corresponds to $T_c < 4 \ {\rm MeV}$ (inaccessible in the case of	
	a standard cosmological history if BBN is to proceed undisrupted if $\alpha_c\sim 1$).  Lines of $\delta =10^{-3}$ and $10^{-2}$ are
	also shown.} 
\end{figure}

\begin{figure}
  \includegraphics[scale=0.53]{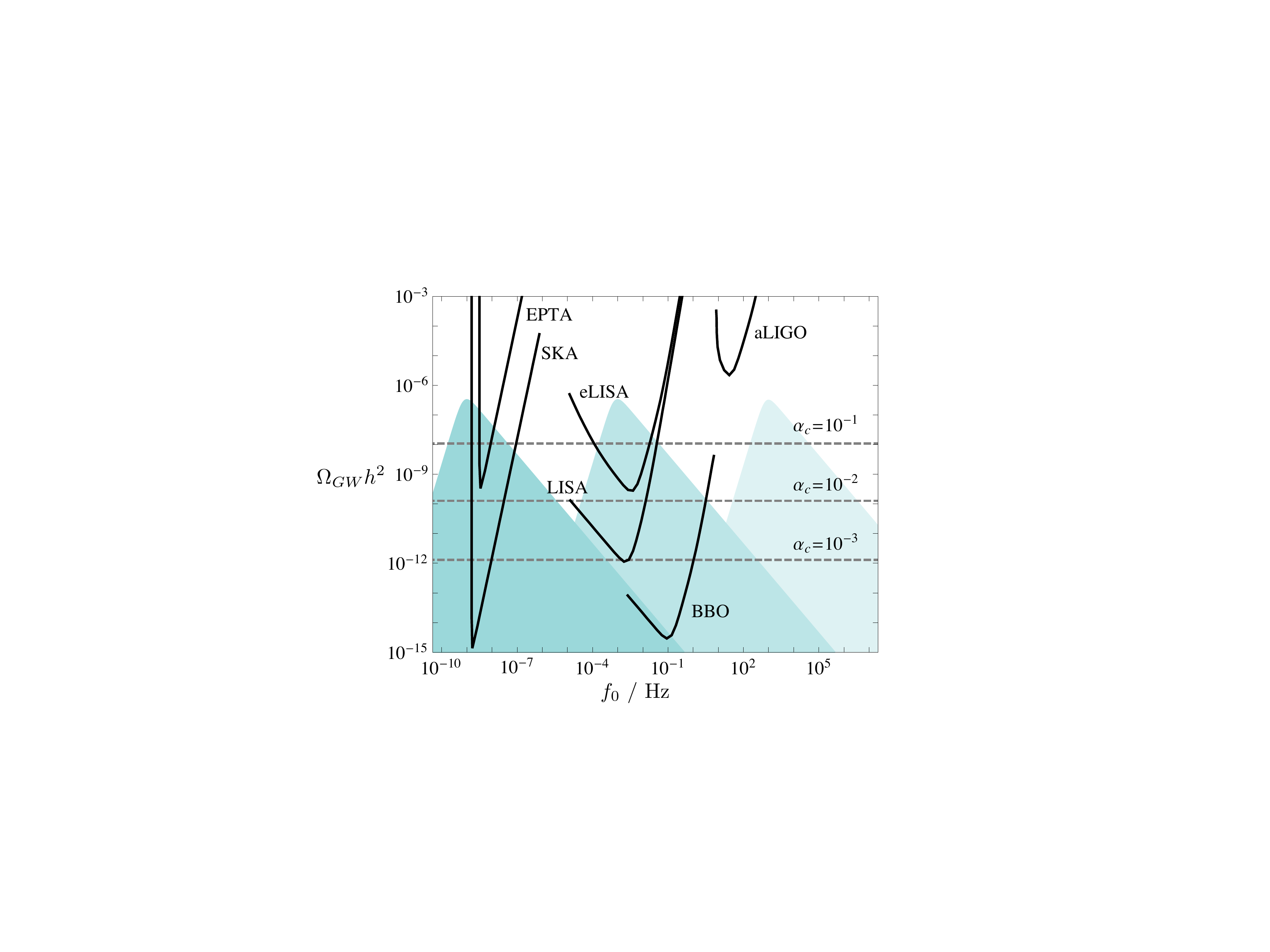}
\caption{\label{fig:GWsignal}
	Approximate profile of the GW signal from bubble collisions, together with the sensitivity curves of different experiments.
	For illustration, we choose 3 different frequency peaks: $f_0 = 10^{-9}, 10^{-3}$ and $10^{3}$ Hz, all within the reasonable range
	of frequencies expected (see Fig.~\ref{fig:freqvsepsilon}), and we set $\alpha_c = 1$.
	Dashed lines show the peak amplitude for other values of $\alpha_c$.} 
\end{figure}

{\bf Collision of bubbles --}
Emission of GWs occurs (predominantly) when the bubbles collide.
Assuming a RD Universe during and after the decay, we can estimate the present frequency peak of the
GW signal as \cite{Leitao:2015ola,astro-ph/9310044}
\begin{equation}
	f_0 \sim 10^{-5} \ {\rm Hz} \ \left( \frac{g_* (T_c)}{100} \right)^{1/6} \left( \frac{T_c}{100 \ {\rm GeV}} \right) \frac{1}{t_* H(T_c)}~,
\label{eq:freqpeak}
\end{equation} 
where $t_*$ refers to the duration of the transition. Given that $\Delta t \sim H(T_n)^{-1}$, we expect $t_* H(T_c) = \mathcal{O}(1)$,
and so we take $t_* H(T_c) \approx 1$ for illustration in Fig.~\ref{fig:freqvsepsilon} and Fig.~\ref{fig:GWsignal}.
(Accurate determination of $t_*$ requires numerical simulations. Note that, although at $T_c$ the dominant collision is between
ultra-relativistic bubbles of size $\sim H(t_n)$, there are also ${\mathcal O}(10-100)$ small, semi-relativistic, thick-walled, bubbles with which
each large bubble collides.)
Fig.~\ref{fig:freqvsepsilon} shows the frequency of the GW signal arising from bubble wall collisions, as a function of the tip warp factor $w_{IR}$.
The frequency of the GW signal can span virtually the whole range of parameter space that will be explored by current and future GW detectors, including 
the high-frequency range, $\gtrsim 30~{\rm kHz}$, where radically new technologies are necessary (but also where conventional astrophysical
foregrounds are absent).
Notice that large values of the volume $\mathcal{V}$ (in units of $(2 \pi l_s)^6$, where $l_s \equiv \sqrt{\alpha^\prime}$)
shift the peak signal frequency towards lower values.
As indicated in the figure, the value of $\delta$ in all cases of interest lies between $10^{-3}$ and $10^{-2}$, well within the range of validity
of our approximation.

Given the unusual nature of the kinetic term of the scalar field $\psi$ (Eq.~(\ref{eq:lagrangian})), a precise expression for the energy density in gravitational radiation observed today, as a function of frequency, would need a dedicated numerical simulation.
However, we believe the \emph{peak} amplitude is well approximated by the usual expression for bubble wall collisions in the case of ultra-relativistic bubbles
\cite{0806.1828,Leitao:2015ola}: 
\begin{equation}
	\Omega_{GW} h^2 (f_0) \sim 10^{-6} \left( \frac{\alpha_c}{1 + \alpha_c} \right)^2 \left( \frac{100}{g_*} \right)^{1/3} \left( t_* H(T_c) \right)^2~.
\label{eq:amplitudepeak}
\end{equation}
In Fig.~\ref{fig:GWsignal}, we show an approximation to the expected signals for three different values of the frequency peak.
Although the peak position should be well approximated by Eq.~(\ref{eq:freqpeak}), the profile of the signal as a function of
frequency beyond the peak requires dedicated numerical simulation. In the present work, we have used
the usual high-$f$ dependence purely for illustration (see \cite{Caprini:2015zlo} for details) . 

\section{IV. Fate of the vacuum energy}
Although enough to produce a potentially observable GW signal, the fraction of the vacuum energy density that converts into gravitational radiation is
small, Eq.~(\ref{eq:amplitudepeak}).
The remaining $\mathcal{O}(1)$ fraction of the false vacuum energy density must therefore have a different fate.
In the following, we outline some of the possibilities.  We stress that this is a very model dependent issue, and unambiguous
statements can only be made on a case-by-case basis.

{\bf Dark radiation -- }
It is possible that some degrees of freedom in the throat will remain massless or very light, even after SUSY breaking.
If some fraction of the vacuum energy density was transferred into these massless states (either directly or indirectly after the decay
of some other massive throat states), they will behave as dark radiation (DR), and constraints from bounds on the number of effective
neutrino species, $\Delta N_{eff}$, will apply.
Since initially $T_{th}\approx 0$, we find that the contribution to $\Delta N_{eff}$ from massless throat states is
\footnote{Notice that $\Delta N_{eff}^{(th)}$ does not depend on the number of massless degrees of freedom in the throat, unlike what would happen
if the SM and the throat sector had been in thermal equilibrium.}
\begin{equation}
	\Delta N_{eff}^{(th)} \approx 0.29 \left( \frac{\eta_{DR}}{1} \right) \left( \frac{\alpha_c}{0.1} \right) \left( \frac{100}{g_*(T_c)} \right)^{1/3}~,
\end{equation}
where $\eta_{DR}$ is the fraction of vacuum energy density that gets transferred into massless states in the throat.

The contribution to $\Delta N_{eff}$ from the SM sector is $\Delta N_{eff}^{(SM)} \approx 0.046$, and Planck has measured $\Delta N_{eff} \approx 0.15 \pm 0.23$ \cite{Ade:2015xua}.
Thus, even for $\eta_{DR} \approx 1$ and $\alpha_c \approx 0.1$, the prediction for $\Delta N_{eff}^{(th)}$ is small enough to be compatible with data
(and an observable GW signal may still be produced, as shown in Fig.~\ref{fig:GWsignal}).
On the other hand, $\alpha_c \approx 1$ would yield too large a value of $\Delta N_{eff}^{(th)}$.

{\bf Black hole production -- }
The large concentration of energy that
takes place when bubbles collide likely leads to the formation of pBHs \cite{PBH1,PBH2,PBH3,PBH5,PBH6}.
Thus, the stringent constraints from energy injection into the thermal plasma due to evaporating pBHs
can apply \cite{Carr:2009jm}.

Depending on the fraction of energy transferred into pBHs of a certain mass $m_{BH}$, certain range of collision temperatures (and therefore of
GW frequencies) could be ruled out, specially for masses $10^9 \ {\rm g} \lesssim m_{BH} \lesssim 10^{18} \ {\rm g}$, where the constraints are strongest.
Determining the extent to which pBH production affects our conclusions would require a detailed numerical simulation, taking into account
the details of the bubble wall profiles at collision.
We note that the production of pBHs with a mass $10^{17} - 10^{18} \ {\rm g}$ accounting for the DM is a possibility,
although its feasibility depends of the fraction of energy density transferred into pBHs.
Formation of pBHs with masses $\sim 30 M_\odot$ (the other possibly open DM window \cite{Bird:2016dcv}) does not seem possible if the Universe remains RD during and after the transition, but less standard cosmological scenarios may allow for this possibility.
We return to this in future work \cite{future1}.

{\bf Non-pBH dark matter -- }
If sufficiently stable states exist in the throat sector, and suitable fraction of $\rho_{vac}$ is transferred to these, they could account for the DM in the Universe, or a component of it.  Although DM candidates arising from warped throats have been studied \cite{Chen:2006ni,Harling:2008px,Frey:2009qb}, and they are indeed a very natural possibility,
whether a given throat can accommodate a DM candidate depends sensitively on the details of the sector living on the throat, as well
as on other aspects such as how SUSY is broken in the rest of the manifold.
Due to how model dependent this issue is, we cannot be more concrete regarding the possibility of successful generation of DM in the scenarios we consider,
or regarding potential constraints from DM overproduction.
However, we note that this is an interesting possibility for DM model building that would lead to a DM candidate with features linked to those of the GW signal discussed in Section~III~\cite{future1}.\\
 
{\bf Final word -- }
GW detection experiments will help shape the future of physics in the coming century.
We believe that they provide an exciting opportunity for the investigation of fundamental physics linked to
the highest energy scales.\\

{\bf Acknowledgments --} 
We would like to thank Joe Conlon and John Wheater for useful discussions.
IGG is financially supported by the STFC/EPSRC and a Scatcherd European Scholarship from the University of Oxford.
SK is supported by the European Research Council under starting grant `Supersymmetry Breaking in String Theory' (307605).
JMR thanks Sidney Coleman for his extraordinary guidance and friendship.


\bibliography{kpvRefs} 

\end{document}